\documentclass[%
reprint,
superscriptaddress,
amsmath,amssymb,
aps,pre,
]{revtex4-2}

\usepackage{graphicx}
\usepackage{dcolumn}
\usepackage[caption=false]{subfig}
\usepackage{bm}
\usepackage{float}
\usepackage{mathtools,amsmath,amssymb}
\usepackage{derivative}
\usepackage{color} 
\usepackage{soul}
\usepackage{xr}
\makeatletter
\newcommand*{\addFileDependency}[1]{
\typeout{(#1)}
\@addtofilelist{#1}
\IfFileExists{#1}{}{\typeout{No file #1.}}
}
\makeatother

\usepackage{hyperref}
\preprint{}

\begin{document}

\title{Structurally balanced growing network as randomized P\'olya urn process
}
\author{Krishnadas Mohandas}
\thanks{Authors contributed equally}
\affiliation{Faculty of Physics, Warsaw University of Technology, Koszykowa 75, PL 00-662 Warsaw, Poland}

\author{Piotr J. G{\'{o}}rski}
\thanks{Authors contributed equally}
\affiliation{Faculty of Physics, Warsaw University of Technology, Koszykowa 75, PL 00-662 Warsaw, Poland}

\author{Krzysztof Suchecki}
	\affiliation{Faculty of Physics, Warsaw University of Technology, Koszykowa 75, PL 00-662 Warsaw, Poland}
\author{Georges Andres}
\affiliation{Chair of System Design, ETH Z\"urich}
\author{Giacomo Vaccario}
\affiliation{Chair of System Design, ETH Z\"urich}
\affiliation{Chair of Ecosystem Magagment, ETH Z\"urich}
\author{Janusz A. Ho{\l}yst}
\affiliation{Faculty of Physics, Warsaw University of Technology, Koszykowa 75, PL 00-662 Warsaw, Poland}

	
\begin{abstract}
    We investigate a process of growth of a signed network that strictly adheres to Heider structural balance rules, resulting in two opposing, growing factions. New agents make contact with a random existing agent and join one of the factions with the bias $p$ towards the group they made contact with. The evolution of the group sizes can be mapped to a randomized P\'olya urn model.
    Aside from $p=1$, the relative sizes of the two factions always tend towards $1/2$, but the behavior differs in the anti-bias regime ($p<1/2$) and the biased one ($p>1/2$).
    In the anti-bias regime, the expected faction sizes converge toward equality, regardless of initial differences, while in the biased regime, initial size difference persists over time. This difference is obscured by fluctuations, with the faction size distribution remaining unimodal even above $p>1/2$, up until a characteristic point $p^{ch}$, where it becomes bimodal, with initially larger and smaller factions featuring their own distinguishable peaks. 
    We discuss several approaches to estimate this characteristic value. 
    At $p=1$, differences between the relative sizes of factions can persist indefinitely, although still subject to fluctuations.  
\end{abstract}
	
\maketitle
	
	
\section{Introduction}
\label{sec:introduction}

Polarization in social networks arises when interactions among individuals lead to the emergence of mutually opposed factions \cite{bramson2016disambiguation,liu2023emergence,guerra2013measure}. 
While many models attribute this phenomenon to a combination of social and structural mechanisms \cite{siedlecki2016,baumann2020modeling,phillips2023high} such as structural balance, homophily, and tribalism \cite{dandekar2013biased,talaga2023polarization,asikainen2020cumulative}, less attention has been paid to how polarization unfolds as networks grow and new members integrate into existing divisions. 
Here, we focus on this growth process and ask how a newcomer’s first, potentially biased interaction can shape the long-term organization of the system. 
We propose a stochastic model that captures the collective dynamics of faction formation under structural balance, where each new agent’s initial friendly or hostile encounter propagates through the network and determines its subsequent alignment. 
This framework reveals how early biases, encoded in a single attachment parameter $p$, can drive the system toward parity, dominance, or persistent bimodality in faction sizes.

Although our approach isolates the growth-driven origin of polarization, it conceptually connects to several, often intertwined mechanisms.
Homophily, the preference for associating with similar others, fosters tightly connected communities of like-minded individuals, giving rise to echo chambers that amplify shared beliefs and mutual reinforcement \cite{axelrod1997dissemination}. 
Structural balance theory provides a complementary perspective, describing how triadic relations stabilize through rules such as “the friend of my friend is my friend” and “the enemy of my friend is my enemy,” leading to the segregation of networks into internally cohesive and mutually antagonistic factions \cite{harary,Malarz2022}.
Strong, positive relationships are maintained within each faction, while negative ties dominate between them.
The interaction between homophily and structural balance can amplify or suppress polarization depending on the underlying social context \cite{Gorski2020,Pham2022}.
Tribalism further reinforces these divisions through emotional commitment and categorical in-group loyalty, heightening group salience and resistance to change \cite{whitt2021tribalism}.
Together, these processes explain why polarized states are both common and persistent, yet they offer limited insight into how polarization emerges and evolves dynamically as systems expand and newcomers form their first social ties.

The mechanisms discussed above—homophily, structural balance, and tribalism—differ in detail but share a common outcome: they generate factions through feedback between local interactions and global structure. 
What remains less clear is how such processes unfold when networks grow and newcomers integrate into existing divisions. 
Empirical studies suggest that initial impressions can have long-lasting consequences, with individuals often maintaining their early alignments over time \cite{szell_multirelational_2010}. Motivated by these observations, we develop an abstract growth framework that captures the essential collective dynamics of polarization without committing to a specific microscopic mechanism. 
Our approach combines two fundamental stochastic principles: reinforcement, represented by the path-dependent P\'olya urn process, and randomness, represented by time-invariant Bernoulli trials. 
The P\'olya component reflects how early interactions or positions can self-reinforce, leading to preferential attachment and entrenched divisions \cite{Collevecchio20131219,he2023social}, whereas the Bernoulli component introduces unbiased stochasticity through a constant probability 
$p$ of a positive first contact. 
Together, these elements provide a minimal yet general framework for understanding the emergence, persistence, and relative dominance of opposing factions during network growth.

In this work, we show that the dynamics of polarization in growing factions can be interpreted equivalently through the lenses of structural balance theory and P\'olya-type reinforcement. 
Although these processes are conceptually distinct, they lead to the same outcome—the emergence and growth of polarized groups. 
Recent studies have employed P\'olya urn frameworks to describe network dynamics \cite{Lu2023hyper,marcaccioli2019polya}, but typically without incorporating structural balance or explicit network growth. 
A related model proposed in Ref. \cite{he2023social} combines these principles to capture how link signs evolve under social reinforcement in fixed-size networks, but does not address the dynamics of expanding systems.

Here, we extend the generalized P\'olya urn process in a randomized version to a growth model that captures the influence of a probabilistic attachment bias $p$ on faction formation and dominance. 
The system consists of two mutually antagonistic factions. 
When a new node (agent) enters, it connects positively to one existing member with probability $p$ or negatively with probability $1-p$, and its subsequent relations are determined by structural balance. 
Varying $p$ governs the collective evolution of faction sizes: as $p$ increases, the system undergoes a transition from a unimodal to a bimodal distribution, reflecting the emergence of two distinct macroscopic outcomes.

Bimodality is a key signature of coexistence across many areas of physics, from statistical and soft matter systems to astrophysical and nuclear contexts \cite{KBinder_1987,chandler1987introduction,ashman1994detecting}. 
Several statistical criteria can be used to assess bimodality \cite{wyszomirski1992detecting,kang2019development,hartigan1985dip,wang2009bimodality,ashman1994detecting}; among them, excess kurtosis provides a convenient though partial indicator, as bimodal distributions often exhibit negative kurtosis due to a flattened central region \cite{freeman2013assessing}. 
In our model, bimodality emerges for $p>p^{ch}$, with a characteristic threshold $p^{ch}\approx 0.836$, beyond which the two faction sizes become clearly distinguishable. 
For $p<p^{ch}$, this difference remains obscured by stochastic fluctuations, which we further analyze through a connection to a randomized P\'olya process. 
This threshold marks the point where early stochasticity gives way to macroscopic order, revealing how small biases in local interactions can crystallize into enduring polarization at the global scale.

In the next section, we systematically explore the faction-growth driven by the attachment bias $p$. 
Subsequently, Sections~\ref{sec:mean evolving cliques} and \ref{sec: distribution of  evolving cliques} present the mathematical formulation for computing the mean sizes of the factions, introducing the rate and master equations. 
Sec.~\ref{sec:bimodality} proposes an approximate criterion to observe bimodality and investigates the characteristic value of bias above which bimodality can be observed.
Sec. \ref{sec:polya} draws parallels between the proposed network growth model and the randomized P\'olya urn process, showing the impact of high bias on faction evolution. Finally, Sec.~\ref{sec:conclusion} summarizes our findings and presents concluding remarks.

\section{\label{sec:model}Model}

Consider two distinct groups, such as rival gangs or political parties. 
A newcomer (e.g., a new resident in a neighborhood) randomly encounters a member of one of the groups. 
The likelihood of meeting a particular group member is proportional to the current size of that group. 
Upon this encounter, the newcomer interacts positively with the met group member with probability $p$ or negatively with probability $1-p$. 
In the case of a positive interaction, the newcomer forms positive ties with all existing members of the group and, at the same time, develops negative ties with all members of the opposing group.

Formalizing the above process leads to considering the following growth dynamics for an undirected signed network. 
In this network, nodes (or agents) represent individuals and links represent relations between them: a positive link $x_{i,j}=1$ indicates a friendly relation between nodes $i$ and $j$, and a negative link $x_{i,j}=-1$ signifies hostility.
At time $t=t_0$, the networks contains $t_0$ nodes labeled $i=1, 2, \ldots, t_0$. 
The network is fully connected and structurally balanced, i.e., the nodes can be split into two factions. 
All the links inside the factions are positive, whereas the links between the factions are negative. 
Then, at every time step $t>t_0$, a new node with label $i=t$ arrives and links to a node $j<t$ chosen uniformly at random from the network.
The sign of the new link is positive with probability $p\in[0,1]$ and negative otherwise.
The probability $p$ serves as an attachment bias and governs the likelihood that a new node forms a positive link (friendly relation) with its first contact.
After establishing its first link, the new node connects with all other existing nodes in the network.
The signs of these links are such that all newly formed triads in the network are structurally balanced. 
In other words, if the new node has a positive link with the randomly sampled node $j$, it joins the $j$ faction and forms positive links with $j$'s friends and negative links with $j$'s enemies. 
Conversely, if the new node has a negative link with the randomly sampled node $j$, it joins the opposing faction of $j$ and the signs of the new connections are reversed.
As a consequence, the growing network is always a complete graph because the newly added nodes connect to all existing nodes. 
Secondly, since every new triad in the network is balanced, the complete graph stays organized into two mutually hostile factions, with the nodes being added to one faction or the other.

When the bias parameter is $p=0.5$, the process is purely random. 
At each step, both factions have an equal probability of gaining a new node, leading to an unbiased evolution. 
For $p<0.5$, the dynamics becomes \textit{anti-biased} as new agents tend to adopt the opposing stance. 
For $p>0.5$, two regimes can be distinguished: a \textit{low-bias} regime, where new agents are only slightly more likely to join the dominant faction, and \textit{high-bias} regime, where the majority faction is strongly favored. 
Although this distinction may appear subtle at this stage, it will become clearer in the following sections. 

\begin{figure}
    \centering
    \includegraphics[width=\linewidth]{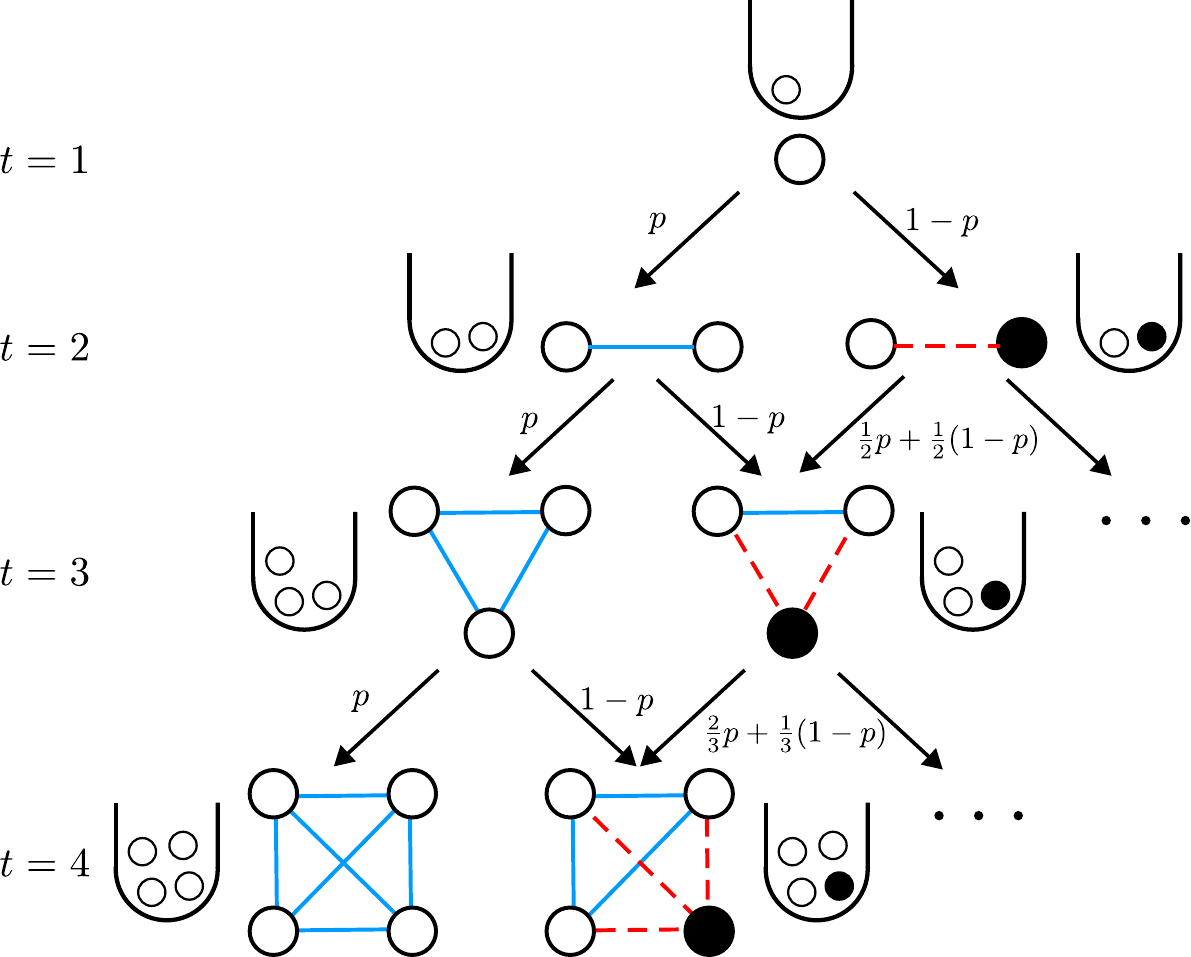}
    \caption{A schematic representation of the growing process can be presented both as a network-based framework and an equivalent urn process. The growth dynamics of the signed network adheres to structural balance and resembles a generalized P\'olya urn process. Here, we start with a single node (white ball). At each time $t$, the generated complete graph organizes into two mutually hostile factions. The solid blue lines correspond to friendly links, and the dashed red lines correspond to hostile links.}
    \label{fig:model}
\end{figure}

Although the model is defined in terms of a signed network, the network structure itself does not play a role in the dynamics; only the sizes of the two factions matter. 
This growth process is similar to a P\'olya urn model, where the number of balls (nodes) in each color (faction) increases according to the reinforcement based on the existing proportions \cite{johnson1977urn, kotz_generalized_2000}. 
Specifically, if we think of one faction as white balls and the other as black balls, then the dynamics can be described as drawing a ball (selecting a node) and adding a new ball of the same color with probability $p$ or of the opposite color with probability $1-p$. 
This attachment bias $p$ also serves as the reinforcement parameter in the analogous P\'olya urn model, controlling the preferential growth of one faction over the other. 
A schematic representation of the growing network, also interpreted through the lens of the P\'olya process, is shown in Fig. \ref{fig:model}.

If the bias parameter is $p = 1$, we have the standard P\'olya urn process.
This corresponds to the case that each newly added node joins the faction of the randomly selected first friend (i.e., the agent $j$). 
In this scenario, the dominant faction, on average, stays dominant, and the stationary distribution of faction sizes follows a beta distribution \cite{johnson1977urn, kotz_generalized_2000}, where the parameters of this distribution are determined by the initial conditions.
If $p \neq 1$, the growth process takes more complex evolutions that we clarify by analyzing the mean faction sizes and their probability distributions.

\section{\label{sec:mean evolving cliques}Mean sizes of evolving factions} 

To understand whether the polarization leads the system to split into two equally-sized groups or to the case when one group is much larger than the other, we must determine the dynamics of the faction-size $m(t)$. 
Let us consider a faction of size $m(t)$ at a time $t$.
The size of this faction can grow when one of the two following events occurs:
\begin{enumerate}
\item a new agent establishes a {\bf positive} link to one of $m(t)$ nodes belonging to this faction.
\item a new agent establishes a {\bf negative} link to one node of the other faction, containing $t-m(t)$ nodes.
\end{enumerate}
The first event occurs with probability $p \frac{m}{t}$, while the second event occurs with probability $(1-p)\frac{t-m}{t}$. 
Since these events are independent, the expected growth of the faction of size $m(t)$ is given by:
\begin{equation}
 m(t+1)-m(t)=p\frac{m}{t}+(1-p)\frac{t-m}{t} \label{eq:delta m}.
\end{equation}
In the continuous limit, the discrete difference $m(t+1)-m(t)$ is approximated by the derivative $\frac{d m}{d t}$, leading to a rate equation (RE) having the following solution:
\begin{equation}
    m(t)=\frac{t}{2}+Ct^{2p-1},
\label{eq: m(t)}
\end{equation}
where $C$ is an integration constant. 
This solution describes the expected size of a faction as the network grows, with the constant $C$ reflecting the influence of the starting configuration.
\begin{figure}
    \subfloat[\label{fig:model_plow}$p=0.3$]{
    \includegraphics[width=\linewidth]{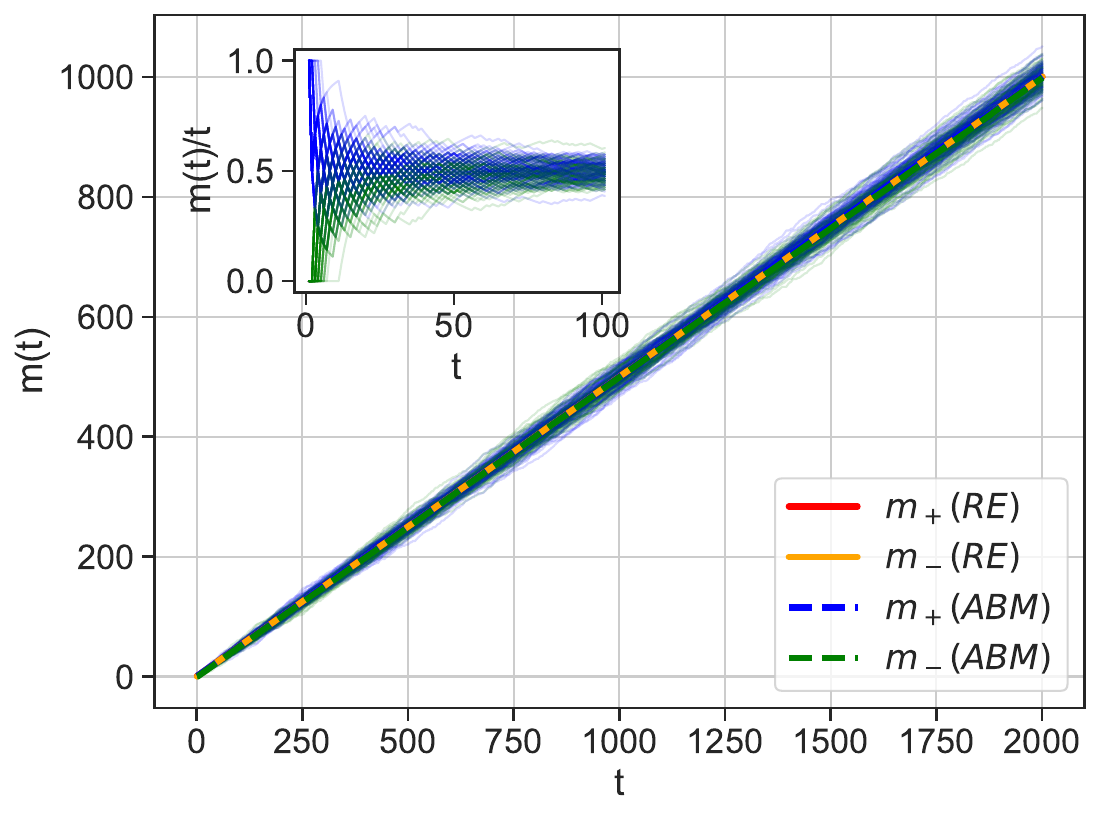}
    }\hfill
    \subfloat[\label{fig:model_phigh}$p=0.9$]{
    \includegraphics[width=\linewidth]{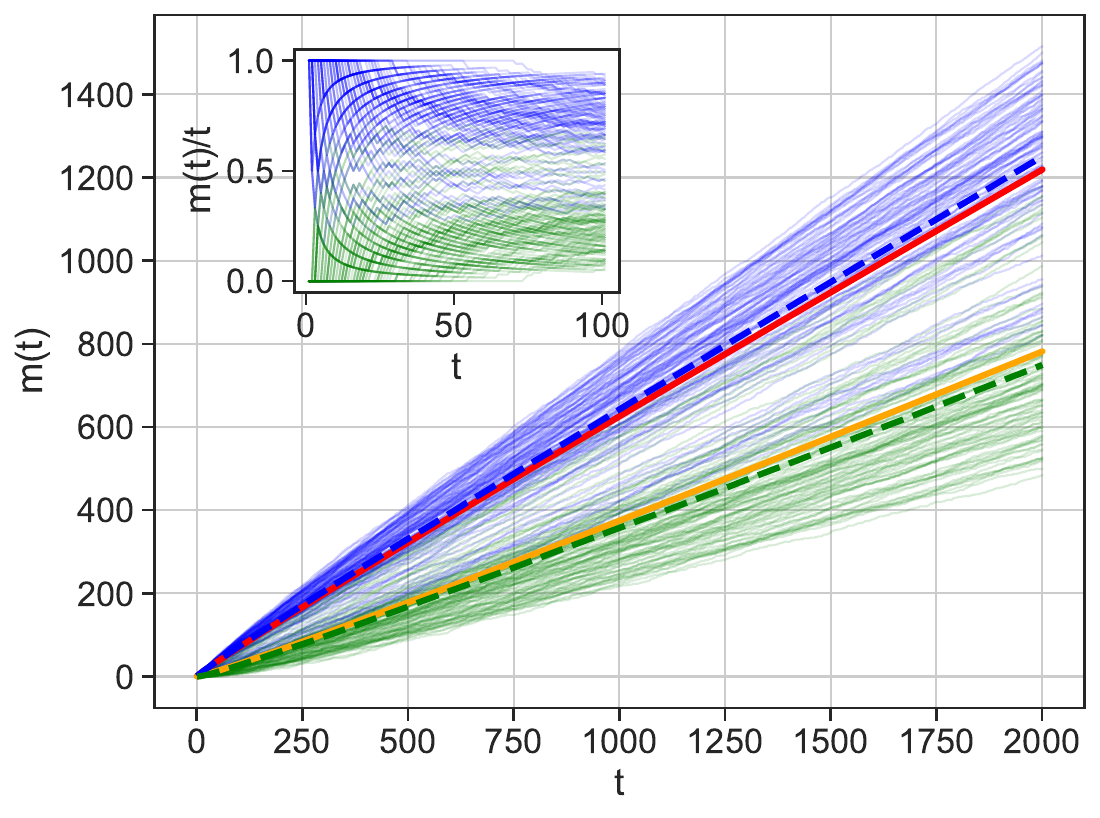}
    }
   \caption{Evolution of sizes of two factions $m_\pm(t)$ for different attachment biases $p$. 
   Expected faction sizes given by the rate equation (RE, solid lines) closely match the respective means of the agent-based model (ABM, dashed lines) simulation. 
   (a) For a small value of $p=0.3$, the mean sizes of the factions are equal. 
   (b) For a large value $p=0.9$, the means $\left< m_\pm(t) \right>$ diverge. Thin lines show individual trajectories of 100 agent-based simulations of the two faction sizes. The initial condition was a single starting node belonging to $m_+$. 
   The inset in each panel shows the corresponding normalized mean over time, which converges over a long time.
   }
    \label{fig:m}
\end{figure}

The value of the integration constant is determined by fixing the initial conditions.
Without loss of generality, consider a network initialized at time $t_0$, consisting of two factions, $m_+(t_0)$ and $m_-(t_0)$, satisfying $m_+(t_0) + m_-(t_0) = t_0$. 
Given these initial conditions, the constant $C$ from Eq.~(\ref{eq: m(t)}) is given by:
\begin{equation}
    m_{\pm}(t_0) = \frac{t_0}{2}+C_\pm t_0^{2p-1} \implies C_\pm = \frac{2m_{\pm}(t_0)-t_0}{2t_0^{2p-1}}.
\end{equation}
Thus, the expected faction size grows as
\begin{equation}\label{eq:general_meanm}
    m(t) = \frac{t}{2}+ \left(m_{\pm}(t_0)-\frac{t_0}{2}\right)\left(\frac{t}{t_0}\right)^{2p-1}
\end{equation}
(see Fig. \ref{fig:m}) and the expected difference in faction sizes is given by:
\begin{align}
    \Delta m(t) \equiv |m_+(t)-m_-(t)|=  \Delta m(t_0) \left(\frac{t}{t_0}\right)^{2p-1}\label{eq:delta_m_RE},
\end{align}
where $\Delta m(t_0)=|m_+(t_0)-m_-(t_0)|$ is the initial difference in faction sizes.

Another way to estimate the mean sizes of factions exploits the connection between the presented stochastic process and the P\'olya urn model.
In the standard P\'olya process the color of the ball drawn from the urn determines the color of the newly added ball. This classical model has been generalized in various ways. One particularly relevant generalization is described as follows \cite{janson_functional_2004}.
After selecting a ball of color $i$, a set of additional balls $[c_1^i, c_2^i]$ is returned to the urn, where $c_j^i$ denotes the number of balls of color $j$ added in response to drawing the color $i$.
In \cite{janson_functional_2004}, the numbers $c_j^i$ are integers sampled from a given distribution.
We instead assume abstractly a continuous (\textit{fractional}) growth version of the process, in which at each step, \textit{fractional} numbers of balls (nodes), $p$ and $1-p$, are added to the respective factions. 
We refer to this variant as the \textit{fractional growth process}. 
The process corresponds to the P\'olya process when the distribution of $c_j^i$ takes two possible values, either $c_j^i=0$ or $c_j^i=1$, with mean equal to $p$.
Under this assumption, with a single node initially, the expected mean faction sizes evolve as (see SM, Sec. III 
for details):
\begin{equation}
    m_{\pm}(t)=\frac{t}{2}\pm\frac{t^{2p-1}}{2\Gamma(2p)},
\label{eq:-m-pm-polya}
\end{equation}
where $\Gamma$ is the gamma function. 

\begin{figure}[!htbp]
    \includegraphics[width=\linewidth]{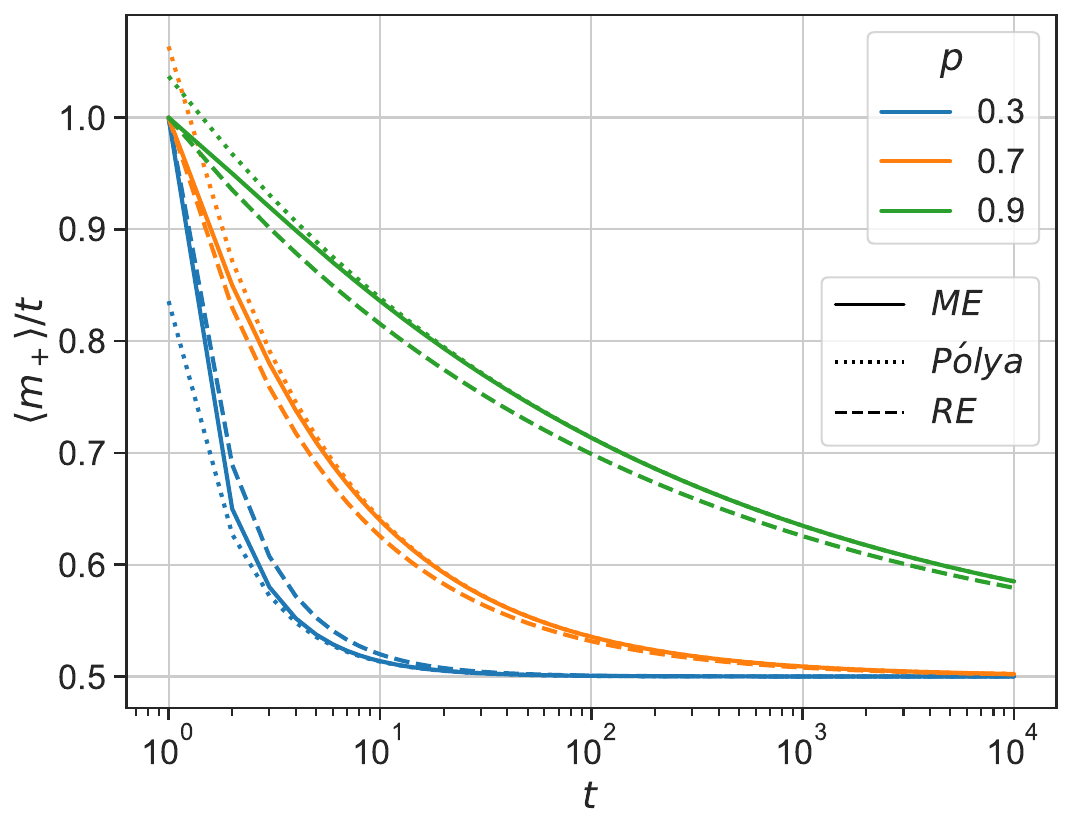}
   \caption{
   In time, the normalized group sizes approach equality. 
   Convergence rates depend on the bias $p$. The system starts at $t_0=1$ with a single node belonging to group $m_+$.
   Exact results derived from the master equation (ME) (\ref{eq:meq}) are shown as solid lines, while approximate values from the P\'olya process formula (\ref{eq:-m-pm-polya}) and the rate equation (RE) (\ref{eq:general_meanm}) (dashed) are shown as dotted and dashed lines, respectively.  
   }
    \label{fig:mean_size}
\end{figure}

This second estimate of the mean faction size better approximates mid-term and long-term behavior compared to the rate equation.
Fig. \ref{fig:mean_size} illustrates this comparison by showing the normalized mean ($\left< {m}_{+}\right >/t$, computed using Eq. \eqref{eq:-m-pm-polya},  Eq.(\eqref{eq: m(t)}, and the exact mean obtained from the master equation \eqref{eq:meq} (derived in the following section).
For anti-bias values of $p$, i.e., when $p<0.5$, the mean rapidly converges to $0.5$, whereas for larger $p$, it approaches this value asymptotically.
The rate equation aligns with the master equation at short timescales, whereas the fractional growth P\'olya process better approximates long-term behavior.
It is worth noting, however, that Eq. (\ref{eq: m(t)}) gives a general solution for any initial distribution of nodes (see SM, Fig. S5)
, while Eq. (\ref{eq:-m-pm-polya}) applies specifically to the case of a single initial node.

Both expressions for $\Delta m(t)$---that in \eqref{eq:delta_m_RE} and the one derived from \eqref{eq:-m-pm-polya}---exhibit the same scaling, $\Delta m\propto t^{2p-1}$. This leads to two interesting observations about the role of the control parameter $p$.
First, for $p<0.5$, $\Delta m$ decreases to zero as the system size $t$ increases (see Fig. \ref{fig:model_plow}), whereas for $p>0.5$, it diverges as $t$ grows (see Fig. \ref{fig:model_phigh}).
This indicates that $p=0.5$ constitutes the first critical point governing the system behavior. 
Second, for all biases $p<1$, the normalized value of this difference, $\delta m=\Delta m/t$, tends to 0 as $t^{2p-2}$ when $t\rightarrow\infty$. 
This can also be seen in Fig. \ref{fig:mean_size}, where all expected normalized means converge universally to $0.5$.
This result has one very interesting implication: as the system size grows, it becomes harder and harder to determine which faction is the dominating one.
To explore the origin of this phenomenon, we next examine the \textit{probability distributions} of the faction sizes and their evolution. 

\section{\label{sec: distribution of evolving cliques}Size distributions of evolving factions} 

The results in Eqs. \eqref{eq:general_meanm} and \eqref{eq:-m-pm-polya} provide analytical expressions for the expected sizes of both factions. 
However, since the system evolves stochastically, due to both the random attachment of new nodes and the process governed by the bias parameter $p$, 
it is necessary to examine its evolution in greater detail by analyzing
the probability distributions of faction sizes. 
Let $P(m,t+1)$ be the probability distribution that a faction is of size $m$ at time $t+1$. 
To calculate this distribution, we consider all possible ways a faction can reach the size $m$ at $t+1$, given the distribution at the previous step $t$. 
This can occur through two mutually exclusive events: (i) a new node is added to a faction of size $m-1$ or (ii) the faction was already of size $m$ and no new node is added to it. 
The former event, i.e., a transition from $m-1$ to $m$, occurs with a rate $w(m \mid m-1)$, while the latter with a rate $w(m\mid m)$.
Then, the master equation (ME) for the process gives the probability $P(m, t)$:
\begin{equation}\label{eq:meq}
    \begin{aligned}
    P(m,t+1) &= P(m,t) +   w(m\mid m-1) \cdot P(m-1,t) \\
    &\quad -\, w(m+1\mid m) \cdot P(m,t), 
\end{aligned}
\end{equation}
where transition rates are as follows:
\begin{equation}
\begin{aligned}
    w(m\mid m-1) &= \frac{(m-1)(2p-1)}{t}+1-p, \\
    w(m+1\mid m) &= \frac{m(2p-1)}{t}+1-p.
\end{aligned}
\end{equation}

The master equation \eqref{eq:meq} characterizes the evolution of the probability distribution, which is inherently non-stationary because of the growth process. 
The shape of this distribution depends on the parameter $p$ and the initial conditions. 
In the following, we examine how different initial conditions give rise to either unimodal or bimodal probability distributions.
Specifically, we distinguish between \textbf{asymmetric} initial conditions, where two starting factions differ in size, and \textbf{symmetric} conditions, where the factions are statistically of equal size. 

\begin{figure*}[htb]
    \subfloat[$p=0.7$\label{fig:dist_p0.7}]{
    \includegraphics[width=0.32\linewidth]{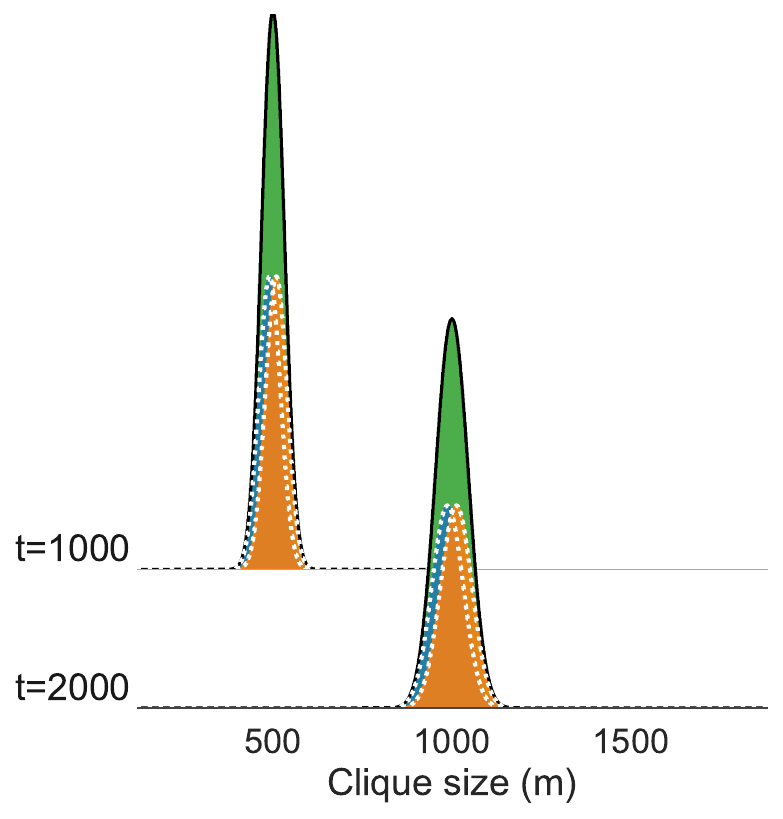}
    
    }\hfill
    \subfloat[$p=0.8$\label{fig:dist_p0.8}]{
    \includegraphics[width=0.32\linewidth]{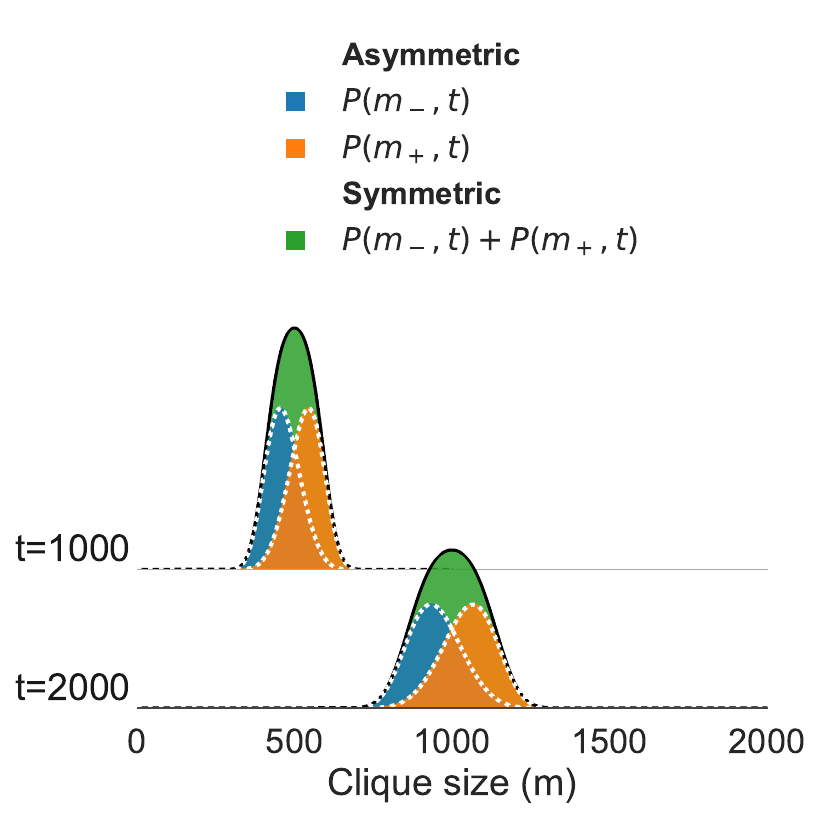}
    }\hfill
    \subfloat[$p=0.9$\label{fig:dist_p0.9}]{
    \includegraphics[width=0.32\linewidth]{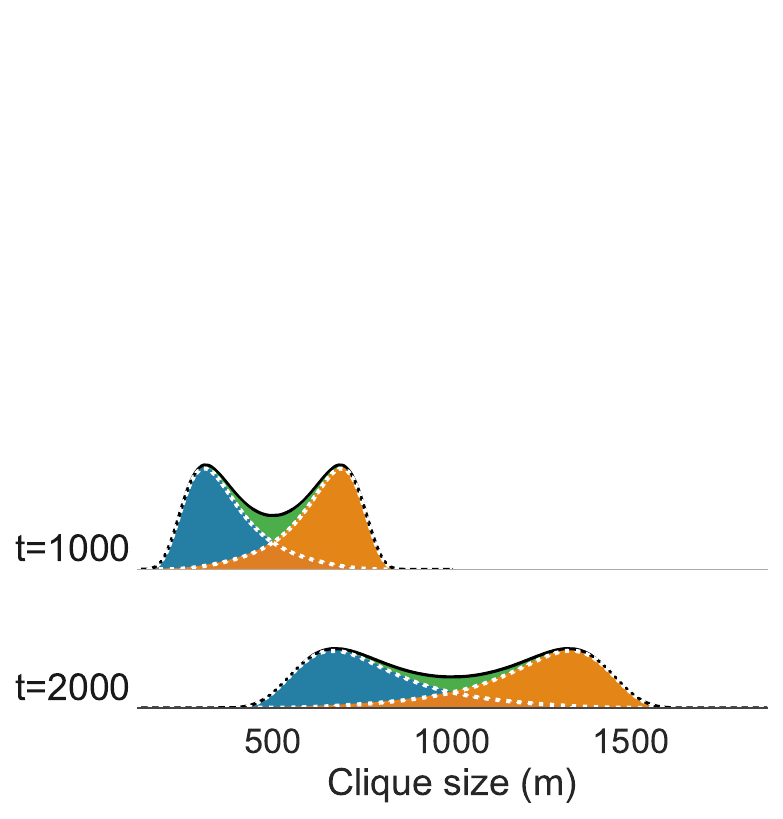}
    }
   \caption{
   Evolution of the three faction size distributions for three values of the attachment bias parameter $p$. 
   Each panel shows the exact distributions derived from the master equation, starting from a single-node system. 
   Two distributions (with dotted outlines) in each panel correspond to an asymmetric initial condition, where the starting node belongs to the faction $m_+$, while the other group $m_-$ is empty. 
   The blue and orange fill are used for the resulting distributions for the initially smaller faction $P(m_-,t\mid m_-^{t_0=1}=0)$ and the larger one $P(m_+,t\mid m_+^{t_0=1}=1)$, respectively.
   Solid outline with green fill represents the distribution under a symmetric initial condition, where the first node is equally likely to join any faction. 
   It also reflects the case where no specific group is observed, and one asks for the probability that any faction is of size $m$.
   That is why it is presented as $P(m_-,t) + P(m_+,t)$. 
   (a) At $p=0.7$, all distributions are unimodal and approximately Gaussian.
   (b) At $p=0.8$, although the asymmetric distributions diverge, the combined result remains unimodal, and no bimodal behavior is observed.
   (c) At $p=0.9$,  the symmetric distribution becomes distinctly bimodal, with the two peaks gradually diverging as new nodes are added.
   }
    \label{fig:model_dist}
\end{figure*}

We first consider the following asymmetric initial condition: $P(m,t=1)=\delta_{m,1}$, meaning that the process begins with a single node in one faction, while the other faction is empty.
We denote the initially larger faction as $m_+$ and the coexisting smaller one as $m_-$. 
Under this initial condition, the resulting distributions $P(m_+,t)$ and $P(m_-,t)=P(t-m_+,t)$ remain unimodal for all values of $p$ (see dashed distribution in orange and blue in Fig.~\ref{fig:model_dist}). 

Second, we consider the symmetric initial condition $P(m,t=1)=\frac{1}{2}\delta_{m,1} + \frac{1}{2}\delta_{m,0}$.
This setup represents a system with two distinguishable factions, where the first node belongs to one of them, chosen uniformly at random. 
In such a case, while we observe one particular faction, the identity of the initially dominant group is unknown.
Although one of the factions starts larger, we do not label it as $m_+$ because its identity is random. To illustrate, we may imagine two factions, ``black’’ and ``white’’: the first agent belongs to either with equal probability. Consequently, the resulting distribution is symmetric around $t/2$. 
As the system evolves, this uncertainty can lead to a bimodal distribution, with each peak corresponding to a different dominant outcome. 
For sufficiently large values of $p$, these two peaks are clearly separated (see Fig. \ref{fig:model_dist}).

An alternative interpretation of the symmetric initial condition is that it represents the probability of observing a faction of size $m$ in a system that began from asymmetric initial conditions. 
In other words, if we start from the asymmetric setup, allow the system to evolve, and then select a faction at random (black or white), the probability that it has size $m$ is given by the symmetric distribution. 
Knowing that one group evolves from $P(m,t=1)=\delta_{m,1}$ and the other from $P(m,t=1)=\delta_{m,0}$, the distribution obtained by randomly choosing either faction corresponds precisely to the symmetric case defined above. 
Thus, the distribution resulting from symmetric initial conditions is directly related to the sum of the two distributions obtained under asymmetric conditions (see legend in Fig. \ref{fig:model_dist}). 
Both frameworks are equivalent since the master equation is linear in $p$. 

By comparing the faction-size distributions at t=1000 and t=2000 (Fig. \ref{fig:model_dist} top vs bottom), we can observe the steady growth of the faction with new nodes continuously added.
Specifically, for the asymmetric initial condition (in blue and orange), the peaks shift to the right for larger $t$, i.e., the mode of the distributions increases over time.
More interestingly, the overlap of the distributions decreases over time, which indicates that one faction grows faster than the other.
In the symmetric case (in green), this distribution appears to be unimodal for smaller $p$ values and becomes evidently bimodal for larger values.
For smaller $p$, the peaks shift to the right but are not distinguishable, see Fig. \ref{fig:model_dist}(a) and (b). 
For larger $p$, the two peaks are evident and shift apart over time, see Fig. \ref{fig:model_dist}(c). 

We further characterize $P(m,t)$ in the SM. 
We compute the skewness, i.e., the third moment of the distribution, and kurtosis, i.e., the fourth moment.
The former allows us to quantify the asymmetry of a distribution, while the latter can be used to estimate its bimodality. 
A description of these measures, along with the corresponding results, is presented in the SM, Fig. S2 and S3.

Based on the derived results, we have two counterintuitive findings.
First, for asymmetric initial conditions, the difference in the mean size of the factions calculated from Eq. \ref{eq:delta_m_RE} increases over time as soon as  $p>0.5$. 
This suggests that the mean sizes of smaller and larger factions diverge as $p$ exceeds this threshold, potentially fostering a bimodal distribution. 
However, the normalized difference between mean sizes goes to zero for all $p<1$.
Second, for symmetric initial conditions, the distribution (absolute or relative) is expected to be bimodal for $p>0.5$.
However, it is not observed until $p$ becomes very large, as seen in Fig. \ref{fig:model_dist}.  
To understand these apparent discrepancies, we need to find a criterion to observe bimodality and study the relation between expected faction sizes and fluctuations.

\section{Criterion for bimodality}
\label{sec:bimodality} 
To quantify the emergence of bimodality and determine the threshold value of the bias parameter at which distinct peaks appear,
we apply the criterion proposed by \cite{Behboodian01021970} for the mixture of two Gaussian distributions:
\begin{equation}\label{eq:bimodalrule}
    |\mu_1 - \mu_2|>2 ~\text{min}(\sigma_1, \sigma_2),
\end{equation}
where $\mu_1$ and $\mu_2$ are the means and $\sigma_1$ and $\sigma_2$ are the respective standard deviations of the Gaussian distributions. 
While this exact criterion does not apply directly in our case, two similar principles hold: i) the bimodal distribution $P(m,t)$ in the analyzed stochastic process can be written as the mixture of two unimodal distributions (see SM for details), and ii) larger separation between means relative to their variances makes the bimodality visible. 

For $p=0.5$, our growth process reduces to a sequence of Bernoulli trials, equivalent to an unbiased random walk.  In this case, the standard deviation $\sigma$ scales as $\sqrt{t}$. 
Although for $p \neq 0.5$, the process becomes biased, we assume the variance follows the same scaling. 
Hence, by analogy with random-walk or Poisson-like fluctuations, the criterion for bimodality \eqref{eq:bimodalrule} can be rewritten as
\begin{equation}
    t^{(2p-1)}  > 2\sqrt{t}  \Leftrightarrow p > \frac{3}{4} + \frac{\log2}{2\log t} .
    \label{eq:cond:std}
\end{equation}
For example, for $t=10^3$, $p>0.75+\log(2)/6\simeq0.800$. 
For $t=2\times10^3$, $p\gtrsim0.75+\log(2)/6.6\simeq0.796$ (See SM, Fig. S9).
These thresholds are consistent with the results shown in Fig.~\ref{fig:model_dist}, where bimodality is observed only for $p=0.9$ as we simulate the growth process up to $t=2\times10^3$.
For large $t$, ($t \to \infty$), analytical threshold of bias parameter estimated from \eqref{eq:cond:std} approaches $\tfrac{3}{4}$. 

\begin{figure}
    \centering
    \includegraphics[width=\linewidth]{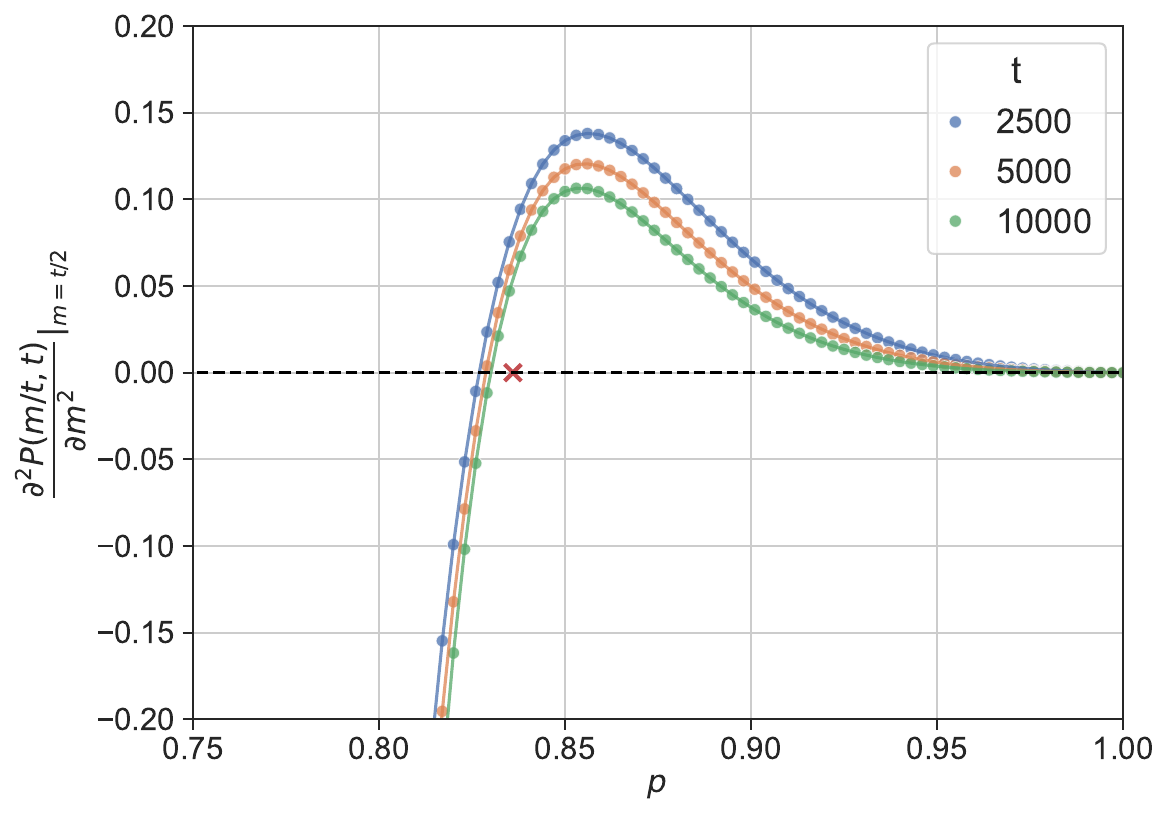}
    \caption{
    Observing bimodality.
    The second derivative of the distribution evaluated at $m = t/2$ serves as an indicator of bimodality; a positive value signifies the presence of a bimodal distribution. For symmetric initial conditions, bimodality persists in the limit $t \rightarrow \infty$ when the attachment bias $p$ exceeds the characteristic value $p^{ch}\approx 0.836$ indicated by a red cross.}
    \label{fig:second_derivative}
\end{figure}

Let us now evaluate this analytical finding with numerical exact results. 
Numerically, a simple and approximate criterion for evaluating the modality of a probability distribution derived from symmetric initial conditions involves examining the change in concavity between the two potential local maxima (modes). This can be achieved by computing the second derivative of the distribution at the midpoint between these maxima. 
A bimodal distribution typically has two distinct peaks separated by a local minimum. At this midpoint, where the distribution transitions between modes, the second derivative is positive if a local minimum exists, indicating convex behavior. By confirming a positive second derivative at the midpoint, one can infer the presence of two distinct peaks, thus indicating bimodality.
Fig. \ref{fig:second_derivative} shows the values of the second derivative at the midpoint, where the bimodality is observed for $p \gtrsim 0.836$.
The point at which the second derivative crosses zero $p_t^{ch}$ shifts slightly over time $t$, i.e., over the system size. By assuming a power-law relation of the form $\log(p_t^{ch}-p_\infty^{ch})\sim \log t$, we estimated the characteristic point as $p_\infty^{ch}\approx 0.836$ (see SM, 
Fig. S8 for details).
Hence, this bias value represents a characteristic threshold that must be exceeded for the bimodality to persist in the limit $t \rightarrow \infty$ when starting from a single-node symmetric initial condition. 

The analytically obtained threshold $p=0.75$ does not represent a true critical point for the onset of bimodality, as it underestimates the transition observed in the system. 
This discrepancy naturally raises the question of its origin. 
In the analytical approach, fluctuations played a central role, obscuring bimodality in the low-bias regime. 
Therefore, in the following section, we examine their influence in greater depth.

\section{Fluctuations around mean sizes of factions as a generalized P\'olya process}
\label{sec:polya}
Following the approach of \citep{janson_functional_2004} for $p\le0.75$ and of the fractional growth process described in Sec. \ref{sec:mean evolving cliques} for $p>0.75$, we can estimate the variance of the process (see SM for the derivation):
\begin{equation}
    \sigma^2 (t) = \begin{cases}
\displaystyle \frac{t}{4(3 - 4p)}, & \text{for } p < 0.75, \\[10pt]
\displaystyle \frac{t \log t}{4}, & \text{for } p = 0.75, \\[10pt]
\displaystyle \frac{t^{2(2p - 1)}}{2}  A, & \text{for } p > 0.75.
\end{cases}
\end{equation}
where $A=\frac{1}{\Gamma(4p - 1)} \cdot \frac{4p^2 - 2}{4p - 3} - \left( \frac{1}{\Gamma(2p)} \right)^2 $.
Note that the estimates for $p<0.75$ and $p>0.75$ should not be used close to $p=0.75$.
These estimates diverge at $p=0.75$ while the simulated values are continuous and very close to the analytical one computed for this specific $p$ (see SM, Fig. S6).

Using these estimates for the variance, we can compute how the signal-to-noise ratio, i.e., $\Delta m(t)/\sigma (t)$, evolves over time:
\begin{equation}
    \displaystyle \frac{\Delta m(t)}{\sigma (t)} \sim \begin{cases}
    t^{2p-1.5} , & \text{for } p < 0.75, \\[10pt]
    \displaystyle \frac{1}{\sqrt{\log t}} , & \text{for } p = 0.75, \\[10pt]
    \displaystyle B, & \text{for } p > 0.75. \\[10pt]
    \end{cases}
\end{equation}
where $B = 2/(\Gamma(2p) A)$, and $\Delta_m$ was obtained from \eqref{eq:-m-pm-polya}. 
From these scaling relations, we find that for $p < 0.75$, the signal-to-noise ratio goes to zero for $t\to \infty$ as $2p-1.5<0$.
This indicates that for $p < 0.75$, it becomes harder and harder to detect any difference in the faction sizes as the system grows.
For $p = 0.75$ a similar finding holds, but the decrease in the signal-to-noise ratio is very slow.
For $p>0.75$, the signal-to-noise ratio is a constant, indicating that the difference in the faction sizes are more stable and detectable.

\begin{figure}
    \centering
    \includegraphics[width=\linewidth]{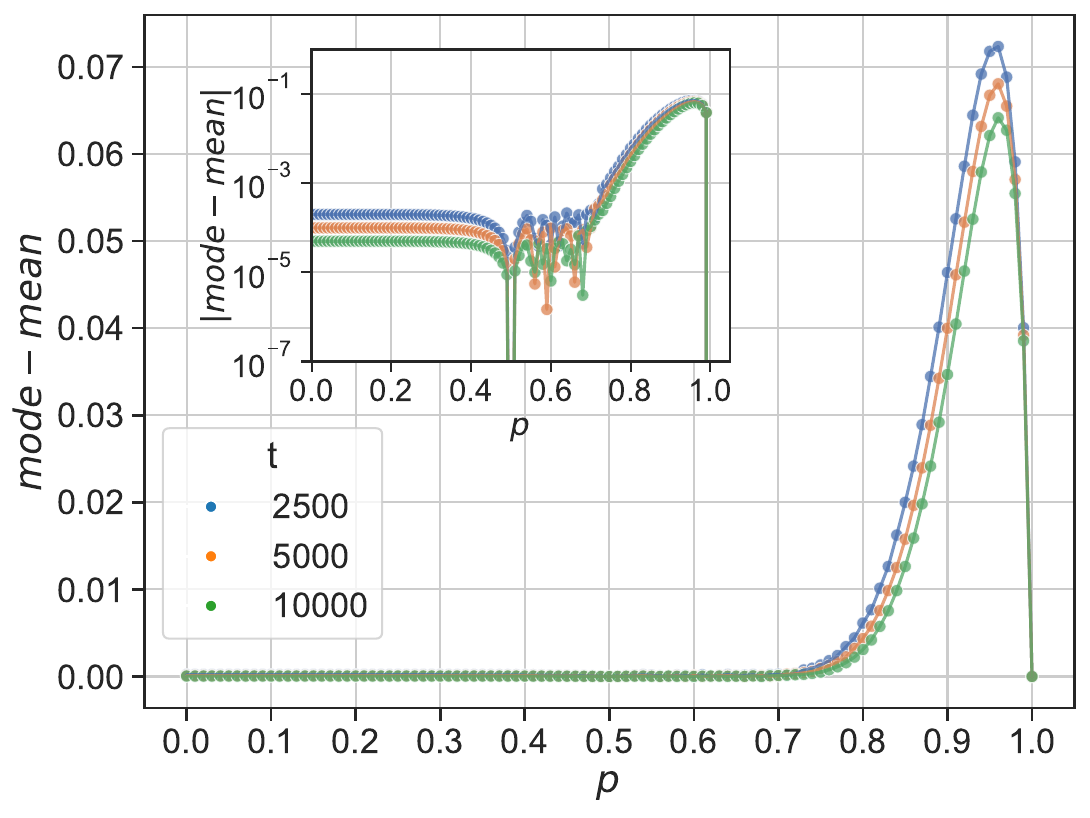}
    \caption{
    Distribution asymmetry is a signature of the high $p$ regime.  
    The figure shows the difference between the normalized mode and the normalized mean, with the inset presenting its absolute value on a logarithmic scale. 
    For $p\gtrsim0.75$, the difference becomes significantly larger than 0, signaling the beginning of the high-bias regime.
    All three distinct regimes of $p$ are clearly evident in the inset. 
    For $p<0.5$, the normalized mode-mean difference behaves smoothly, remaining close to $1/2t$ due to the discrete nature of the distribution, whereas for $0.5<p<0.75$, it exhibits fluctuations. 
    These trends are approximately independent of time.
    }
    \label{fig:mode_mean}
\end{figure}

These analytical expectations are confirmed by simulations (see SM, Fig. S7)
and have one interesting consequence.
In Figure \ref{fig:mode_mean} we show the normalized difference between the distribution's mode and mean.
Recalling that the normalized mean converges to 0.5 (see Fig.  \ref{fig:mean_size}), this difference tells us how bigger (or smaller) is the dominating (dominated) faction.
This difference remains negligible for $p \lesssim 0.75$, then increases sharply, peaking near $p \approx 0.95$, before falling back to zero at $p = 1$, where all new nodes join the single, starting one.
Moreover, in the anti-bias regime $p < 0.5$, the difference is theoretically expected to approach zero, but, due to finite-size effects, it remains slightly greater than zero. 
Although the mean and mode appear almost equal below $p\approx 0.75$, the logarithm scale (see inset of Fig. \ref{fig:mode_mean}) reveals another change of behavior between anti-bias and low-bias values of $p$. 
For $p<0.5$, the asymmetry behaves smoothly, remaining close to
$1/2t$ due to the discrete nature of the distribution, whereas fluctuations without any clear trend occur within the range $0.5 < p < 0.75$.

Recall, however, that to observe bimodality under symmetric initial conditions, the control parameter must satisfy $p>p^{\textrm{ch}}=0.836>0.75$. 
For asymmetric initial conditions, bimodality does not occur. Instead, for $p>0.75$, the mode, which is the most probable size, exceeds the mean, representing the expected size. 
Taken together, these observations suggest that $p \approx 0.75$ does not mark the onset of bimodality, but rather signals a gradual change in the system’s behavior.

\section{Discussion and Conclusion}\label{sec:conclusion}
We have introduced a stochastic growth model describing the formation of two antagonistic factions under structural balance.
At each time step, a new agent joins one of the factions through two sequential mechanisms: a P\'olya urn process and a Bernoulli trial with probability $p$.
The P\'olya urn process favors attachment to the larger faction, establishing a path-dependent growth dynamic; the subsequent Bernoulli trial introduces size- and time-independent randomness that determines whether the agent aligns with the P\'olya bias (with probability $p$) or reverts it (anti-bias, with probability $1-p$).
Analytical expectations for the evolution of faction sizes are derived using the master equation, the rate equation, and the Pólya process formalism.
All approaches yield consistent results, in agreement with numerical simulations, providing a coherent theoretical understanding of the system’s behavior.

Three regimes summarize the system’s behavior. 
For $p<0.5$, anti-bias drives rapid equalization of group sizes. 
The dynamics reflect contrarian behavior \cite{GALAM2004453}, causing the system to reinforce the minority group.
In the low bias regime $0.5<p<p^{ch}$, mean faction sizes diverge, but fluctuations mask this asymmetry, keeping the distribution unimodal. 
For the high bias regime $p>p^{ch}$, where $p^{ch} \approx 0.836$, bimodality emerges, reflecting stable coexistence of unequal factions. 
A notable crossover near $p\approx 0.75$ corresponds to a change in fluctuation scaling rather than a true phase transition.
The transition from unimodal to bimodal distributions resembles the emergence of spontaneous population segregation studied, for example, in \cite{weidlich1991synergetics}. 
A similar transition has been observed in broader models of social dynamics \cite{Weidlich1994267}, where the parameter $p$ corresponds to the strength of herding pressure. 

Another interesting takeaway is that the normalized faction sizes always equalize unless $p=1$. 
This means that even a small probability of joining the opposing side prevents one group from achieving permanent dominance. 
Consider a scenario in which an uninformed individual encounters information online from one side of a debate (for example, online reviews favoring one brand of smartphone over another).
If there is even a slight chance that this individual is not persuaded and instead ends up joining the opposing camp ($p<1$), then the two opposing camps will tend toward the same size, without one or the other winning.
This contrasts with traditional opinion-dynamics models, where conversion processes amplify majorities.
In our model, opinions do not convert opponents but rather compete to attract newcomers, i.e., those who have yet to form an opinion.
Integrating our growth-based mechanism with such models—e.g., voter or majority-rule frameworks \cite{holley1975ergodic,galam2002minority}—could provide a unified view combining internal opinion change with external recruitment.
Finally, relaxing the assumptions of complete connectivity or perfect balance—by allowing probabilistic linking or sign noise as in \cite{li2016model}—may yield richer dynamics, including the emergence of multiple antagonistic clusters or metastable imbalances.

The attachment bias $p$ represents the tendency of newcomers to align with or oppose the first group they encounter. 
It summarizes, in a simplified way, various individual-level influences that might affect group choice—such as access to resources, perceived reputation, or social cohesion—without modeling them explicitly \cite{poepsel2013joining,simpson_cumulative_2017,hohman2021why}. 
In this sense, $p$ acts as an aggregate parameter that captures how local structure and interactions shape the overall balance between the two factions. 
The emergence of size inequality for $p>p^{ch}$ is qualitatively similar to the disparities observed in models of productivity-based or reputation-driven group competition \cite{simpson_cumulative_2017,perez2022emergence}.

Traditionally, network science has emphasized local structural features, such as clustering, modularity, and structural balance measures, as primary explanatory variables for the emergence and persistence of polarized groups~\cite{newman2006modularity,doreian1996partitioning,fortunato2010community}.
These approaches typically center on individuals whose specific decisions based on local information shape the formation of network connection patterns, thereby generating a divided system. 
In contrast, our model shows that stable group divisions can emerge from a simple, group-level stochastic process, without requiring detailed assumptions about individual decision-making.

Our approach highlights the importance of stochastic processes in shaping social and network dynamics, which were observed in other studies \cite{pham2021balance, stadtfeld2020emergence} to explain the emergence of groups. 
The combination of individual and group perspectives has also been explored in more complex models involving hypergraphs, where interactions extend beyond pairwise links to group-level relations \cite{majhi2022dynamics}.
In this sense, our model can be interpreted as a limiting case of hypergraph-based models that capture polarization and group formation \cite{xu2023linear,landry2023opinion,pister2024stochastic}.

Summing up, this work advances our understanding of how structural balance principles interact with randomized attachment dynamics, which might offer insights into the self-organization of growing social alliances, adversarial communities, and other polarized systems.

\begin{acknowledgements}
This research received funding from the Polish National Science Center under Alphorn Grant No. 2019/01/Y/ST2/00058. This work was funded by the European Union under the Horizon Europe grant OMINO (Grant No. 101086321) and was also co-financed with funds from the Polish Ministry of Education and Science under the program entitled International Co-Financed Projects. Research of J.A.H. was also funded by Warsaw University of Technology within the Excellence Initiative: Research University (IDUB) programme and by a special grant for leaders of European projects coordinated by the Warsaw University of Technology.
\end{acknowledgements}

\bibliographystyle{apsrev4-2}
\bibliography{apssamp}

\end{document}